\documentclass[10 pt]{article}
\usepackage{latexsym}
\usepackage{amsmath}
\usepackage{epsfig}

\def\be{\begin{equation}}
\def\ee{\end{equation}}
\def\bea{\begin{eqnarray}}
\def\eea{\end{eqnarray}}
\def\Z{$\mathrm{Z}$}
\def\W{$\mathrm{W}$}

\def\MZ{$\mathrm{M}_{\mathrm{Z}}$}
\def\MW{$\mathrm{M}_{\mathrm{W}}$}
\def\MH{$\mathrm{M}_{\mathrm{H}}$}
\def\VJK{$\mathrm{V}_{\mathrm{jk}}$}
\def\mt{$\mathrm{m}_{\mathrm{t}}$}
\def\GM{$\mathrm{G}_{\mathrm{\mu}}$}
\def\al{$\alpha$}

\begin{document}

\baselineskip = 18pt

\author{
Arif Akhundov $^{a, b}$ \\
\vspace{.5 cm}\\
$^a$ Institute of Physics, Azerbaijan Academy of Sciences,\\
       H. Cavid ave. 33, 370143 Baku, Azerbaijan\\
$^b$ Departamento de F\'isica Te\'orica and IFIC, Universidad de
Valencia-CSIC,\\
       46100 Burjassot (Valencia), Spain
}

\title{Precision Tests of Electroweak Interactions \footnote{Plenary talk at UAE-CERN Workshop: High Energy Physics and Applications,
Al-Ain, UAE, 26-28 Nov. 2007.}}

\maketitle

\begin{abstract}
 The status of the precision tests of the electroweak interactions is
 reviewed in this paper. An emphasis is put on the Standard Model 
 analysis based on measurements at LEP/SLC and the Tevatron. 
 The results of the measurements of the electroweak mixing  
 angle in the NuTeV experiment and the future prospects are discussed.
 
\end{abstract}


\section{Introduction}
  The unification of the electromagnetic and weak interactions in 
1968~\cite{WS:68}, the discoveries of the neutral currents in 
1973~\cite{NC:73}, of the charm quark in 1974~\cite{CQ:74}, of 
the \W~and \Z~bosons in 1983~\cite{WZ:83} were very successful 
  steps for the theory of the ElectroWeak (EW) interactions, the 
  Standard Model (SM)~\cite{WS:68,GIM}.  
  After the discoveries of the top quark in 1995~\cite{CDFD0:95} and 
  the tau neutrino in 2000~\cite{TN:00} the electroweak SM became the commonly 
  accepted theory of the fundamental electroweak interactions. It is a 
  gauge invariant quantum field theory based on the symmetry group 
  $\mathrm {SU(2)\times U(1)}$,  which is spontaneously broken by the Higgs mechanism. The renormalizability
  of the SM~\cite{TH:71} allows us to make precise predictions for  
  measurable quantities at higher orders of the perturbative  
  expansion, in terms of a few input parameters. The higher-order terms,
  Radiative Corrections (RC) or quantum corrections, contain the 
 self-coupling of the vector bosons as well their interactions with the 
  Higgs field and the top quark. Their calculation provides the 
  theoretical basis for the EW precision tests. 

  In the last thirty five years in High Energy Physics two distinct 
  and  
  complementary strategies have been used for gaining new understanding
  of the Nature:
\begin{itemize}
\item The Direct Discovery of the New Phenomena at High Energy  
      accelerators
\item The Precision Measurements of the Known Phenomena at existing  
      accelerators with high Luminosity
\end{itemize}

  The excellent example is the ratio (Fig. 1)
\noindent 
\bea
\mathrm R_{e^+e^-} \;\equiv\;
{\sigma(e^+e^-\to \mbox{\rm hadrons})\over\sigma(e^+e^-\to\mu^+\mu^-)} \, .
\eea
\noindent 
\begin{figure}
  \includegraphics[height=.3\textheight]{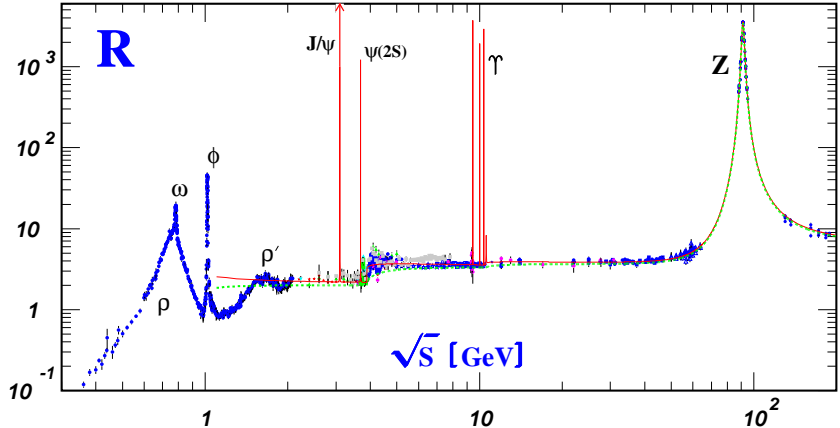}
  \caption{World data on the ratio $\mathrm R_{e^+e^-}$~\cite{PDG}.}
\end{figure}
  Both strategies have worked very well for the studies of 
  the EW interactions first at SPS and LEP1/SLC, and later at LEP2/SLD and 
  the Tevatron, and in the future such interplay of the proton and
  electron colliders will be applied for the LHC and ILC/CLIC.  
  
  The second strategy has always demanded from the theory the  
  prediction of physical quantities with high precision, 
  $\emph i.e.$ at the level of quantum corrections. For the 
  interpretation of the precision experiments the RC play the crucial 
  role.    
\section{Electroweak Radiative Corrections}
In the context of the SM any electroweak process can be computed at tree level from \al~(the fine structure constant measured at values of~
$\mathrm{Q^2}$ close to zero), \MW~(the W-boson mass), 
\MZ~(the Z-boson mass), and \VJK~(the Cabbibo-Kobayashi-Maskawa flavor mixing matrix elements)~\cite{CKM}. 

When higher order corrections are included, any observable can be predicted using the on-shell renormalization scheme~\cite{Sir, Hioki} 
as a function of:
\noindent 
\bea
\mathrm{O_i} &=& \mathrm {f_i}(\alpha, \alpha_{\mathrm{s}},
                  \mathrm{M}_{\mathrm{W}} ,\mathrm{M}_{\mathrm{Z}},
\mathrm{M}_{\mathrm{H}},\mathrm{m}_{\mathrm{f}},\rm{V}_{\rm{jk}}) ,  
\eea
\noindent 
where the effects of heavy particles do not decouple,
and there is the sensitivity to the top mass \mt~\cite{ABR1:86} and to less extend to the Higgs mass \MH~\cite{ABR2:86}.
\begin{figure}
\begin{center}
  \includegraphics[width=10cm]{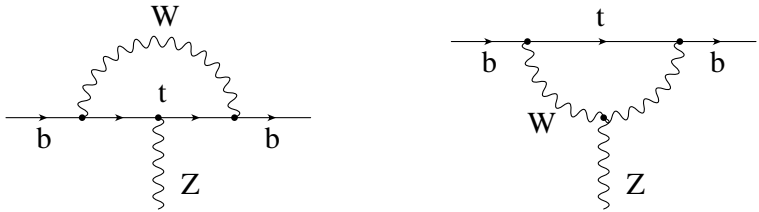}
\end{center}
  \caption{One-loop corrections to the $\mathrm{Z\bar b b}$ vertex, involving a virtual top quark \cite{ABR1:86}.}
\end{figure}
Since the discovery of the presence of hard \mt$^2$ corrections to the 
$\mathrm{Z\bar b b}$  vertex (see Fig. 2)~\cite{ABR1:86,BH:88,LS:90,BPS:91} the calculation of the EW RC has been theoretically well established and many higher-order contributions of
the radiative corrections have become available over past decades to improve and stabilize the SM predictions.
\section{Tests of EW Interactions at LEP/SLC and Tevatron}
The experimental data for testing of the EW theory have achieved an impressive accuracy. After taking the measured \Z~mass, besides
\al~and \GM~(the Fermi constant measured in the muon decay), for completion of the input, 
\noindent 
\bea
\mathrm{M_Z} & = & (91.1875\pm 0.0021)\,\mathrm{GeV}
\quad\mbox{\cite{LEPEWWG,LEPEWWG_SLD:06}}\, ,\\
\mathrm{G}_{\mathrm{\mu}} & = & (1.166\, 371 \pm 0.000\, 006) \cdot 10^{-5}\:\mathrm{GeV}^{-2}
\quad\mbox{\cite{MuLan:07}}\, , \\
\alpha^{-1} & = & 137.035\, 999\, 710\pm 0.000\, 000\, 096
\quad\mbox{\cite{GHKNO:06}}
\eea
\noindent 
each other precision observable provides a test of the electroweak theory (Fig. 3). The predictions are calculated with computer programs
ZFITTER~\cite{ZFITTER} and TOPAZ0~\cite{TOPAZ0}, which incorporate
state-of-the-art calculations of the EW, QED and QCD radiative
corrections.

Theoretical predictions of the SM depend on the mass of the top quark
and of the as yet experimentally unknown Higgs boson through the virtual presence of these particles in the loops. 
As a consequence, precision data can be used to pin down the allowed 
range of the mass parameters. This is shown in Fig. 4, which compares the information on \MW~and \mt~obtained at LEP1 and SLD, with the
direct measurements performed at LEP2 and the Tevatron. 

The measured at Tevatron mass \mt~$= 172.6 \pm 1.4$~\cite{CDFD0:08} 
agrees better than 10 \% with the value predicted within the SM on the basis of the precision EW measurements. 

\begin{figure}
\begin{center}
  \includegraphics[height=.4\textheight]{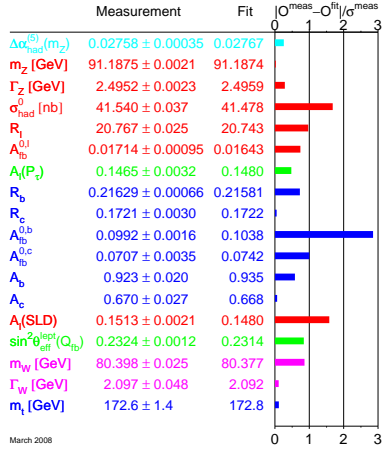}
\end{center}
  \caption{Comparison between the measurements included in the combined analysis of the SM and the results from the global EW fit~\cite {LEPEWWG,LEPEWWG_SLD:06}.}
\end{figure}
\begin{figure}
\begin{center}
  \includegraphics[width=6cm]{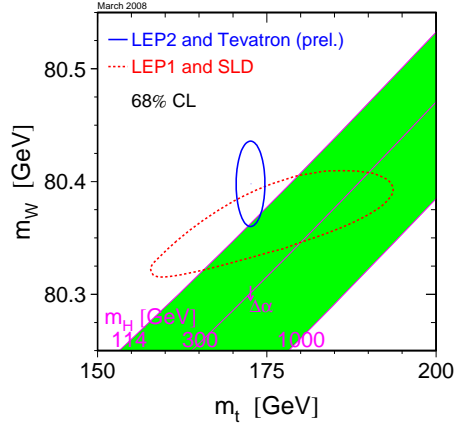}
\end{center}
  \caption{Comparison of the direct measurements of \MW~and
\mt~at LEP2/Tevatron with the indirect determination
through electroweak radiative corrections at LEP1/SLD. Also shown
in the SM relationship for the masses as function of 
\MH~\cite{LEPEWWG, LEPEWWG_SLD:06}.}
\end{figure}

Taking all direct and indirect data into account, one obtains the
pillar of the precision electroweak physics~\cite{Ellis:07}: the
best constraints on the possible mass \MH~of unseen Higgs. The global electroweak fit results in the
$\Delta\chi^2 = \chi^2-\chi^2_{\rm min}$ curve shown in
Fig. 5. The lower limit on \MH~obtained from
direct searches is close to the point of minimum $\chi^2$. At 95\%
C.L., one gets~\cite{LEPEWWG,LEPEWWG_SLD:06}
\noindent 
\bea
114.4\;\mathrm{GeV}\; <\; \mathrm{M_H}\; <\; 160\;\mathrm{GeV}. \nonumber 
\eea
\noindent 

\begin{figure}
\begin{center}
  \includegraphics[width=6cm]{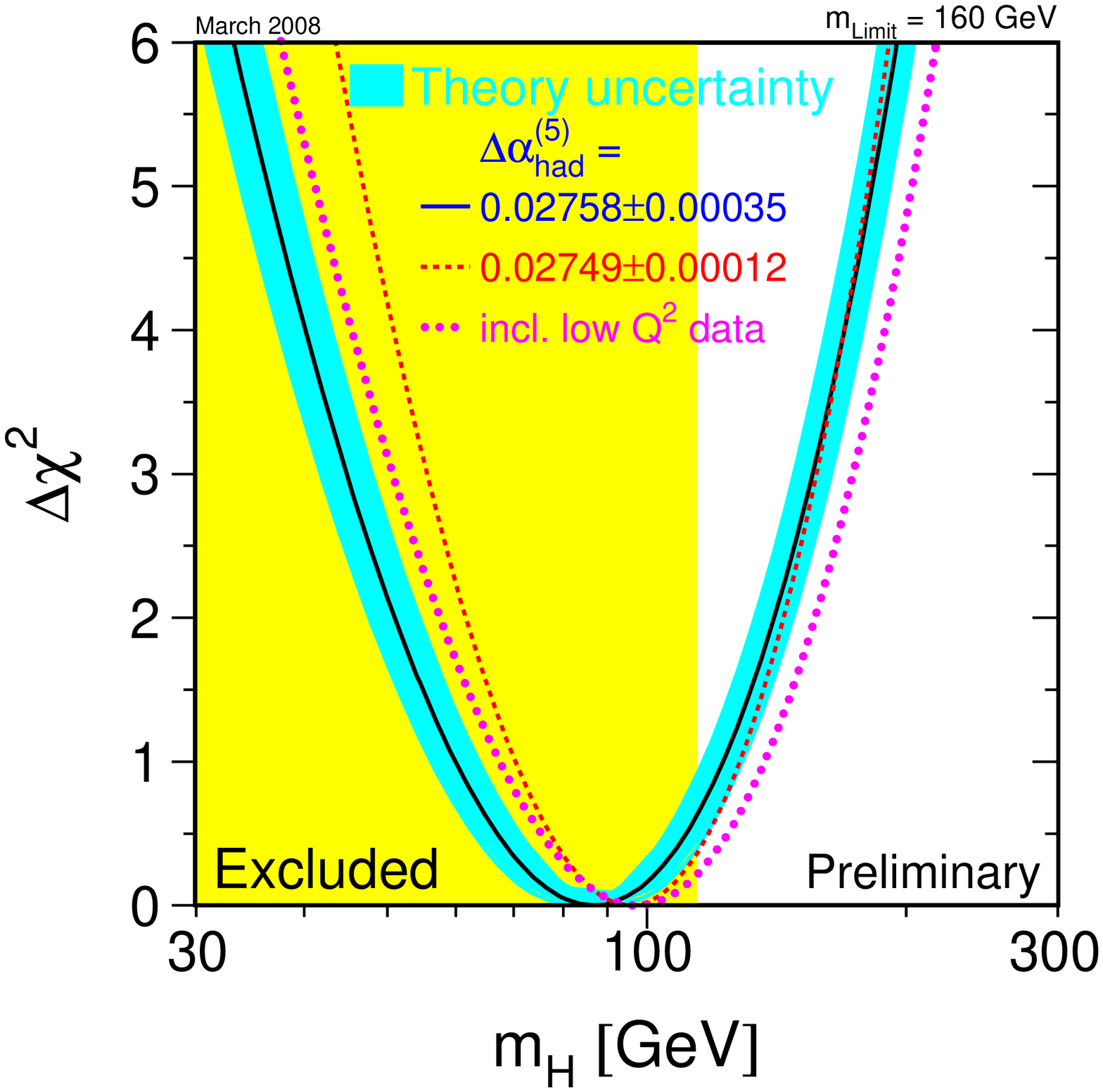}
\end{center}
  \caption{$\Delta \chi^2 = \chi^2-\chi^2_{\rm min}$ versus \MH,
from the global fit to the EW data. The vertical band
indicates the 95\% exclusion limit from direct searches
\cite{LEPEWWG,LEPEWWG_SLD:06}.}
\end{figure}


\section{SM Analysis and NuTeV Experiment}

In the on-shell scheme \cite{Sir, Hioki} the three-level formula
$\sin^2{\theta_W} = 1 - {\mathrm{M}_\mathrm{W}^2\over
\mathrm{M}_\mathrm{Z}^2}$ is a definition of the
renormalized $\sin^2{\theta_W}$ to all orders in perturbation theory,
$\emph i.e.$,  
\noindent 
\bea
\sin^2{\theta_W}^{\mbox{on-shell}} \; \equiv\; {\mathrm{s}_\mathrm{W}}^2 \; = \; 1 - {\mathrm{M}_\mathrm{W}^2\over\mathrm{M}_\mathrm{Z}^2}\, .
\eea
\noindent 
A precise determination of the on-shell EW mixing angle has been performed by the NuTeV collaboration~\cite{NuTeV} for the first time through the measurements of the Pashos-Wolfenstein ratio
\cite{PW:73}:
\noindent 
\bea
\mathrm R^{-} &\equiv& \frac{\sigma(\nu_{\mu}N\rightarrow\nu_{\mu}X)-
                   \sigma(\bar\nu_{\mu}N\rightarrow\bar\nu_{\mu}X)}
                  {\sigma(\nu_{\mu}N\rightarrow\mu^-X)-  
                   \sigma(\bar\nu_{\mu}N\rightarrow\mu^+X)}   
\eea
\noindent 
from deep inelastic neutrino scattering on isoscalar targets.
The NuTeV collaboration finds 
${\mathrm{s}_\mathrm{W}}^2=0.2277 \pm 0.0016$ which is 
3.0 $\sigma$ higher than the SM predictions. 

From this experimental value one obtains the mass of \MW~boson
~\cite{NuTeV}
\noindent 
\bea
\mathrm{M_W} = 80.14 \pm 0.08 \; \mathrm{GeV}\, 
\eea
\noindent 
which is smaller than other measurements of~\MW~at LEP/SLD and the
Tevatron (see Fig. 6).
\begin{figure}
\begin{center}
  \includegraphics[width=6cm]{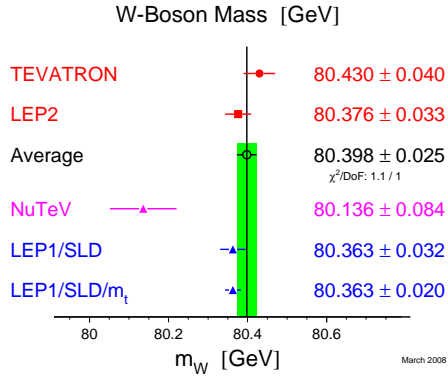}
\end{center}
  \caption{The results of the direct measurements of~\MW~at 
 LEP2/Tevatron are compared with the indirect determinations at LEP1/SLD 
 and in the NuTeV experiment~\cite{LEPEWWG,LEPEWWG_SLD:06,NuTeV}.}
\end{figure}
The NuTeV result should be considered as preliminary until a reanalysis
of data will be completed including all experimental and theoretical
information. 

\section{Conclusions and Future Prospects}

  Apart from the still missing Higgs boson, the SM provides an elegant 
  theoretical framework for the description of the known experimental  
  facts in Particle Physics. The SM has been impressively confirmed by 
  successful collider experiments at the particle accelerators LEP, SLC 
  and Tevatron during the last fifteen years. 
  
  Future colliders like the upcoming LHC or an ILC/CLIC offer great
  prospects, and in turn represent a great challenge for theory
  to provide even more precise calculations. Accurate predictions are
  necessary not only to increase the level of precision of SM tests, but
  also to study the indirect effects of possible new particles.

\section{Acknowledgments}
 I would like to thank the organizers of the UAE-CERN Workshop 
 for kind invitation to give a plenary talk and for the hospitality 
 at Al-Ain. I am grateful to J. Vidal 
 for support and to D. Hertzog, Z. Hioki and A. Shiekh for remarks 
 and comments.



\end{document}